\pdfoutput=1

\documentclass[12pt]{article}
\usepackage{jheppub}

\usepackage{amsmath}
\usepackage{amssymb}
\usepackage{graphicx,color,slashed,braket}
\usepackage{bbm}
\usepackage{pifont}
\usepackage{ifpdf}
\usepackage{comment}
\usepackage{array}
\usepackage{float}

\newcommand{\be}{\begin{equation}}
\newcommand{\ee}{\end{equation}}
\newcommand{\ba}{\begin{eqnarray}}
\newcommand{\ea}{\end{eqnarray}}

\newcommand{\lp}{\left(}
\newcommand{\rp}{\right)}

\newcommand{\e}{\textrm{e}}

\newcommand{\w}{\wedge}

\title{T-dualities and scale-separated AdS$_3$ in type I}

\author{Zheng Miao$^a$, Muthusamy Rajaguru$^{a,b}$, George Tringas$^a$, and Timm Wrase$^a$}

\affiliation{$^a$Department of Physics, Lehigh University, 16 Memorial Drive East, Bethlehem, PA 18018, USA}
\affiliation{$^b$Instituto de Física Téorica IFT-UAM/CSIC, C/ Nicolás Cabrera 13-15, Campus de
Cantoblanco, 28049 Madrid, Spain\\}

\emailAdd{zhm323@lehigh.edu, muthusamy.rajaguru@uam.es, georgios.tringas@lehigh.edu, timm.wrase@lehigh.edu}

\notoc

\abstract{
\noindent
We perform three T-dualities on previously found, classical $\mathcal{N}=1$ scale-separated AdS$_3$ solutions of massive type IIA supergravity. These solutions arose from a compactification on a toroidal $G_2$-holonomy space with smeared O2/D2 and O6/D6 sources.
The T-dual backgrounds are classical $\mathcal{N}=1$ AdS$_3$ solutions of type IIB supergravity with O5/D5 and O9/D9 sources (type I) compactified on a space with $G_2$-structure and non-vanishing Ricci scalar.
We generalize the original solutions in IIA in the T-dual picture and present on the type IIB side fully classical solutions with parametric control, scale separation, and integer conformal dimensions for the dual operators in the corresponding CFT. We also obtain strongly coupled solutions with the same properties. These are S-dual to parametrically controlled classical solutions of the heterotic SO(32) string theory.}

\begin{document}

\vspace*{-2cm} 
\begin{flushright}
{\tt \phantom{xxx} IFT-UAM/CSIC-25-97} \qquad \qquad
\end{flushright}

\maketitle

\newpage

{\hypersetup{hidelinks}
\tableofcontents
}

\section{Introduction}\label{sec:introduction}

In recent years, significant progress has been made in exploring scale-separated solutions within type II supergravity, following a different direction from earlier efforts that primarily focused on constructing four-dimensional theories. Recent studies have instead turned to lower-dimensional models, particularly in three-dimensional AdS constructions \cite{Farakos:2020phe,Emelin:2021gzx,Emelin:2022cac,VanHemelryck:2022ynr,Farakos:2023nms,Farakos:2023wps,Arboleya:2024vnp,VanHemelryck:2025qok,Farakos:2025bwf,Arboleya:2025lwu}, as well as to certain aspects of scale separation in two dimensions \cite{Lust:2020npd,Cribiori:2024jwq} and five dimensions \cite{Cribiori:2023ihv}, and to understanding uplifts to M-theory \cite{Cribiori:2021djm,VanHemelryck:2024bas}.
For a broader study of scale separation across various dimensions, see \cite{Tringas:2025uyg}, and for a comprehensive review see \cite{Coudarchet:2023mfs}.

Until recently, the literature focused on four-dimensional AdS vacua in massive type IIA, particularly the constructions of \cite{DeWolfe:2005uu,Camara:2005dc}, building on earlier general work \cite{Behrndt:2004mj,Derendinger:2004jn,Lust:2004ig}.
Further developments exploring four-dimensional scale-separated vacua in massive IIA have appeared in \cite{Ihl:2006pp,Marchesano:2019hfb,Carrasco:2023hta,Tringas:2023vzn}.
To further understand these four-dimensional constructions, the literature has examined uplifts to ten dimensions \cite{Acharya:2006ne}, challenges related to localization and Romans mass \cite{Saracco:2012wc,DeLuca:2021mcj}, and critical assessments of uplift validity \cite{McOrist:2012yc}. See \cite{Junghans:2020acz} for a perturbative examination of the localization approximation and \cite{Emelin:2022cac} for implications in other dimensions.
Scale-separated AdS constructions also face challenges from the Swampland program \cite{Palti:2019pca}, in particular \cite{Lust:2019zwm} and the conjectural arguments in \cite{Montero:2024qtz} questioning their robustness, as well as from tests of flux-vacuum consistency and holography \cite{Apers:2025pon}.
We note that, in contrast to the three-dimensional case, classical four-dimensional solutions with scale separation in type IIB have not yet been realized, despite early attempts in \cite{Caviezel:2009tu,Petrini:2013ika} and further commentary in \cite{Cribiori:2021djm}.

Going back to the focus of this work, which is three-dimensional constructions, minimal (non)-supersymmetric three-dimensional AdS vacua were constructed from massive type IIA compactifications on isotropic $G_2$-holonomy orbifolds in \cite{Farakos:2020phe} and verified using bispinor methods in \cite{VanHemelryck:2022ynr}, leading to parametric scale separation, parametric control, and stabilized moduli.
This construction exhibited the same features as the DGKT setup, despite differences in the internal space and spacetime dimensions. The emergence of such universal behavior, independent of specific model characteristics, was justified in \cite{Tringas:2025uyg}.
Subsequent works explored anisotropic orbifolds, focusing on scale separation and the study of dual conformal dimensions \cite{Farakos:2023nms,Farakos:2023wps}, with parametric scale separation and integer conformal dimensions explicitly realized in \cite{Farakos:2025bwf}. As will be discussed later in more detail, scale separation was achieved in all these constructions, while the appearance of integer conformal dimensions was tied to the specific mechanism used to cancel the tadpole.

On the other hand, the construction of minimal three-dimensional AdS vacua in type IIB can be achieved using internal spaces with $G_2$-structure and non-vanishing Ricci scalar.
The initial constructions in \cite{Emelin:2021gzx}, later verified via bispinor methods in \cite{VanHemelryck:2022ynr}\footnote{See also \cite{Passias:2019rga,Passias:2020ubv} for AdS$_3$ vacua in type IIB with $G_2$-structure, obtained using the bispinor formalism.}, used a twisted toroidal orbifold as the internal space, taken to be isotropic.
In that setup, since the internal space was isotropic, scale separation is obstructed due to the non-vanishing Ricci scalar \cite{Gautason:2015tig,Tringas:2025uyg}. The only way to decouple the AdS scale, which is proportional to the internal curvature, from the KK scale would be by tuning the metric fluxes, which would violate flux quantization. Nevertheless, the presence of an O9/D9 system, implying that the theory is actually type I, allowed for an exact match with the heterotic construction of \cite{deIaOssa:2019cci} via S-duality, highlighting a nice feature and illustrating the connection between the two theories.
Subsequently, related but non-supersymmetric constructions based on solvmanifolds were studied in~\cite{Arboleya:2024vnp}, while supersymmetric solutions on solvmanifolds and nilmanifolds were explored in \cite{VanHemelryck:2025qok}, where these setups do not include the O9/D9 system.
There, it is shown that nilmanifolds appear to allow for scale-separated vacua but it is unclear whether they exist for compactifications on solvmanifolds. However, for both types of compactification spaces it is possible to find solutions for which the operators dual to the lightest scalars in AdS have integer conformal dimensions.

Apart from the above constructions that use O-planes, recent work has explored scale-separated AdS solutions in six-dimensional gauged supergravity \cite{Proust:2025vmv} and has also pursued approaches based on Casimir energies \cite{Aparici:2025kjj}.
As noted in the previous paragraphs, there is ongoing interest in using the pattern of conformal dimensions as a practical probe of the link between scale-separated AdS vacua and their holographic duals, first discussed in \cite{Apers:2022zjx,Apers:2022tfm,Apers:2022vfp} and further discussed in \cite{Plauschinn:2022ztd}.

In this paper, we perform three T-dualities on classical scale-separated AdS$_3$ solutions of massive type IIA \cite{Farakos:2025bwf} with $G_2$-holonomy, obtaining type I backgrounds with $G_2$-structure. The dualities and source configuration produce O9-planes canceled by D9-branes, yielding a type I background. Unbounded type IIA fluxes map to closed type I fluxes and provide parametric control, while bounded type IIA fluxes map to bounded type I fluxes and ensure tadpole cancellation. Beyond the dual solutions with parametrically small cycles, a flux-scaling analysis yields scale-separated families that are either classical or at strong coupling, the latter dual to heterotic classical solutions~\cite{Het}. The inherited flux configuration stabilizes the moduli and reproduces the type IIA pattern in which the conformal dimensions become parametrically integer.

\section{The setup}

In this section we will review particular flux compactifications of massive type IIA supergravity on spaces with $G_2$-holonomy. These give rise to three dimensional theories that preserve a priori four supercharges. An additional orientifold projection leads to O2/O6-planes and projects out half the supercharges, so that we are left with minimal supergravity. While one can in principle study general $G_2$ manifolds, we restrict to a particular toroidal orbifold that was extensively studied before. We will work in the orbifold limit, studying only bulk moduli and neglecting light modes arising from the twisted sectors. 

Upon performing three T-dualities, the $H_3$ flux on the type IIA side turns into geometric flux on the type IIB side. This means the internal space has now a non-vanishing Ricci scalar and $G_2$-structure. Below we review the geometric details of the particular toroidal orbifold of interest and introduce our notation.

\subsection{\texorpdfstring{A toroidal orbifold with $G_2$-structure}{}}

A particular $G_2$ space is characterized by the following three-form $\Phi$ and it's dual four-form $\Psi$
\begin{align}\label{phipsi}
    \Phi&=e^{127}-e^{347}-e^{567}+e^{136}-e^{235}+e^{145}+e^{246} \,, \\
    \Psi&=e^{3456}-e^{1256}-e^{1234}+e^{2457}-e^{1467}+e^{2367}+e^{1357} \,,
\end{align}
with $e^{127}=e^{1}\wedge e^{2}\wedge e^{7}$ etc., representing wedge products of the vielbeins $e^{i}$, with $i=1,\dots, 7$, on the $G_2$ space.
The deformations of the internal space, which describe the sizes of the three-cycles, are described by seven $s^i\equiv s^i(x^\mu)$ moduli 
\begin{equation}\label{sdef}
    \Phi=\sum_{i=1}^7s^i\Phi_i\,,
\end{equation}
while the three-form basis $\Phi_i$, and the corresponding four-form basis $\Psi_i$, are defined as
\begin{align}
    \Phi_i&=\left(dy^{127},-dy^{347},-dy^{567},dy^{136},-dy^{235},dy^{145},dy^{246}\right)\,,\label{basis} \\
    \Psi_i&=\left(dy^{3456},-dy^{1256},-dy^{1234},dy^{2457},-dy^{1467},dy^{2367},dy^{1357}\right) \,,\label{basis2}
\end{align}
where $i=1,\dots,7$ and $\Phi_1=dy^1 \wedge dy^2 \wedge dy^7 =dy^{127}$, $\Phi_2=-dy^{347}$ etc.\,. The integrals of the basis elements over the $G_2$ space satisfy
\begin{equation}\label{eq:PhiiPsij}
    \int\Phi_i\wedge\Psi_j=\delta_{ij} \,.
\end{equation}
Then, denoting the internal space by $X$, the volume $\text{vol}(X)$ of the $G_2$ manifold can be written as
\begin{equation}\label{volumeG2}
    \text{vol}(X)=\frac{1}{7}\int\Phi\wedge\star\Phi =\left(\prod_{i=1}^7s^i\right)^{1/3}\,,
\end{equation}
while $\int_X dy^{1\dots 7}=1$. Above and in the following the Hodge star $\star$ denotes the Hodge star on the internal $G_2$ only and we will use $\star_{10}$ to denote the Hodge star of the 10D spacetime.

One finds the following Hodge dual expressions, which will be useful later when discussing fluxes threading the relevant cycles
\begin{equation}
    \star\Phi=\sum_i\frac{\text{vol}(X)}{s^i}\Psi_i\,,
    \qquad\qquad
    \star\Phi_i=\frac{\text{vol}(X)}{\left(s^i\right)^2}\Psi_i \,.
\end{equation}

\subsubsection{\texorpdfstring{$G_2$-holonomy on a toroidal orbifold}{}}

We now focus on the simplest toroidal orbifold example, where the internal space is a seven-torus $X = T^7 / (\mathbb{Z}_2\times\mathbb{Z}_2\times\mathbb{Z}_2)$ modded out by the orbifold group $\Gamma$ with three generators, $\Gamma =\langle \Theta_{\alpha},\Theta_{\beta},\Theta_{\gamma}\rangle$. These act on the internal coordinates as follows 
\be
\begin{aligned}\label{Z2s}
\Theta_\alpha : (y^1, \dots, y^7 ) & \to (-y^1, -y^2, -y^3, -y^4, +y^5, +y^6, +y^7) \, , 
\\
\Theta_\beta : (y^1, \dots, y^7 ) & \to (-y^1, -y^2, +y^3, +y^4, -y^5, -y^6, +y^7) \, ,
\\
\Theta_\gamma : (y^1, \dots, y^7 ) & \to (-y^1, +y^2, -y^3, +y^4, -y^5, +y^6, -y^7) \, . 
\end{aligned}
\ee
It is easy to consider products of these generators, e.g., $\Theta_{\alpha} \Theta_{\beta}=\Theta_{\alpha\beta}$, which give rise to additional generators such as $\Theta_{\alpha\beta}, \Theta_{\alpha\gamma}, \Theta_{\beta\gamma}, \Theta_{\alpha\beta\gamma}$, all of which leave the space invariant. While they can be generated straightforwardly, their explicit form is given in \cite{Farakos:2020phe}, along with a more detailed analysis of the material discussed in this section. The upshot is that the only invariant forms on the toroidal orbifold are the 3- and 4-form given in equations \eqref{basis}, \eqref{basis2}, as well as the volume 7-form $dy^{1234567}$.

The vielbeins of the space are specified in terms of the torus radii as $e^i = r^i dy^i$, and with \eqref{phipsi} and \eqref{sdef} the moduli can then be expressed accordingly; for example, $s^1 = r^1 r^2 r^7$, and so on.
Since this toroidal orbifold has $G_2$-holonomy and thus it is a flat space, the differential forms in \eqref{phipsi} are both closed and co-closed
\begin{equation}\label{G2closed}
    d\Phi= d\star\Phi=0 \,.
\end{equation}

For the type IIA constructions in this paper we will consider an additional O2/O6 orientifold projection whose spatial $\mathbb{Z}_2$ involution $\sigma$ reverses the signs of all internal directions
\begin{equation}
    \sigma\,:\,y^i\rightarrow -y^i \quad\quad \text{for}\quad\quad i=1,\dots,7 \,.
\end{equation}
In addition to O2-planes at the fixed point locations of $\sigma$, one finds a web of O6-planes from the combined actions of the involution $\sigma$ and elements of the orbifold group $\Theta_m$ for $m=\alpha, \beta, \gamma, \alpha\beta, \alpha\gamma, \beta\gamma, \alpha\beta\gamma$.
The internal transverse directions (–) and wrapped directions ($\times$) of the respective O-planes are summarized in the following table
\begin{align}\label{OplanesIIA}
\begin{pmatrix} 
&{\rm O}2:\quad & - & - & - & - & - & - & -  \\
&{\rm O}6_{\alpha}:\quad & \times & \times & \times & \times & - & - & -  \\
&{\rm O}6_{\beta} :\quad &\times & \times & - & - & \times & \times & -  \\
&{\rm O}6_{\gamma}:\quad &\times & - & \times & - & \times & -& \times  \\
&{\rm O}6_{\alpha\beta}:\quad & - & - & \times & \times & \times & \times & -  \\ 
&{\rm O}6_{\beta\gamma} :\quad & - & \times & \times & - & - & \times & \times  \\ 
&{\rm O}6_{\gamma\alpha} :\quad &- & \times & - & \times & \times & - & \times  \\ 
&{\rm O}6_{\alpha\beta\gamma}:\quad &\times & - & - & \times & - & \times & \times 
\end{pmatrix} \, . 
\end{align}

\subsubsection{\texorpdfstring{Co-calibrated $G_2$-structure on a twisted toroidal orbifold}{}}

The type IIB compactifications we are interested in correspond to internal spaces with metric fluxes. As a result, the internal space is no longer Ricci-flat, and the deviation from special holonomy is characterized by the presence of non-zero torsion classes. Since our spaces of interest have no 1- and 2-forms, the only non-zero torsion classes are the 0-form $W_1$ and the 4-form $W_{27}$. These are defined through the following exterior derivatives
\begin{equation}\label{G2nonclosed}
    d\hat{\Phi}=W_1\star\hat{\Phi}+W_{27} \,,\quad\quad
    d\star\hat{\Phi}=0 \,,
\end{equation}
where $\hat{\Phi} \wedge W_{27}=0$.

Above we have added a hat on the form $\hat{\Phi}$ to indicate that we replaced the differential forms $dy^i$ with the twisted 1-forms, $dy^i\rightarrow\eta^i$. These satisfy the Maurer-Cartan equations
\begin{align}
    d\eta^i=\frac{1}{2}\tau^i_{jk}\eta^j\wedge\eta^k\,,
\end{align}
where $\tau^i_{jk}$ are the structure constants of the Lie algebra of the group manifold. The structure constants are constrained by the following two conditions
\begin{align}
    \tau^i_{ji}=0\,,\quad\quad \tau^l_{~[ij}\tau^m_{~k]l}=0 \,.
\end{align}
The first condition is automatic if we T-dualize $H_3$ flux and also automatic for compact spaces \cite{Danielsson:2011au}. The second condition is simply the requirement that $d^2 \eta^m=0$.\footnote{From a physics point of view this requirement corresponds to the absence of geometric sources like KK monopoles.}

The twisted vielbeins are simply given by $\hat{e}^i=r^i\eta^i$. We can then again perform orbifolds that act on the $\eta^i$ in the same way as they did on the $y^i$ above in equation \eqref{Z2s}. The invariant 3- and 4-forms are obtained from equations \eqref{basis} and \eqref{basis2} by replacing $dy^i \to \eta^i$ or simply via
\begin{equation}\label{basisdef}
    \Phi_i\rightarrow\hat{\Phi}_i\,,\quad\quad
    \Psi_i\rightarrow\hat{\Psi}_i\,,
\end{equation}
where $i=1,\dots,7$ and $\hat{\Phi}_1=\eta^{127}$, $\hat{\Phi}_2=-d\eta^{347}$ etc.\,.
It is useful to introduce the following general expressions, which we will use extensively
\begin{equation}\label{dPhi}
    d\hat{\Phi}_i=\sum_j\mathcal{M}_{ij}\hat{\Psi}_j \,.
\end{equation}
This notation was first introduced in \cite{DallAgata:2005zlf} and the matrix ${\cal M}_{ij}$ was explicitly presented in \cite{Emelin:2021gzx}. It has the following entries
\be\label{mijmatrix}
{\cal M}_{ij} = \begin{pmatrix}
  0 & -\tau^{7}_{5,6}  & -\tau^{7}_{3,4}  & +\tau^{1}_{4,5}  & +\tau^{2}_{4,6} & +\tau^{1}_{3,6}  & -\tau^2_{3,5} \\ 
  -\tau^{7}_{5,6} & 0 & +\tau^7_{1,2} & +\tau^3_{2,5} & -\tau^3_{1,6} & -\tau^4_{2,6} & -\tau^4_{1,5} \\
  -\tau^{7}_{3,4} & +\tau^7_{1,2} & 0 & +\tau^6_{2,4} & +\tau^5_{1,4} & -\tau^5_{2,3} & +\tau^6_{1,3} \\
  +\tau^1_{4,5} & +\tau^3_{2,5} & +\tau^6_{2,4} & 0 & -\tau^3_{4,7} & +\tau^1_{2,7} & +\tau^6_{5,7} \\
  +\tau^2_{4,6} & -\tau^3_{1,6} & +\tau^5_{1,4} & -\tau^3_{4,7} & 0 & -\tau^5_{6,7} & -\tau^2_{1,7} \\
  +\tau^1_{3,6} & -\tau^4_{2,6} & -\tau^5_{2,3} & +\tau^1_{2,7} & -\tau^5_{6,7} & 0 & +\tau^4_{3,7} \\
  -\tau^2_{3,5} & -\tau^4_{1,5} & +\tau^6_{1,3} & +\tau^6_{5,7} & -\tau^2_{1,7} & +\tau^4_{3,7} & 0 
\end{pmatrix} \, . 
\ee
In type IIA we will consider an internal space that has $G_2$-holonomy, and therefore the internal Ricci scalar in this case is zero. On the type IIB side, however, the internal curvature is non-zero and can be expressed in terms of the torsion classes
\begin{equation}
R^{(7)}= \frac{21}{8} W_1^2 -\frac{1}{2}\vert W_{27}\vert^2  \, . 
\end{equation}
The torsion classes are in turn related to the structure constants $\tau^i_{jk}$ of the internal space via the matrix ${\cal M}_{ij}$. Their explicit form follows from equation \eqref{dPhi}
\begin{equation}
W_1 = \frac{\sum_{ij} s^i {\cal M}_{ij} s^j}{7 \text{vol}(X)} \,, \qquad W_{27} =  \sum_{i,j} \lp s^i {\cal M}_{ij} -\frac{\sum_k s^i {\cal M}_{ik} s^k}{7 s^j} \rp \hat{\Psi}_j \,.
\end{equation}

For the type IIB constructions in this paper we will consider an additional O5/O9 orientifold projection. In particular we will mod out by the worldsheet parity $\Omega_p$, which leads to a spacetime filling O9-plane. We will cancel its charge and tension by 32 D9-branes and therefore have what is usually called type I string theory. However, we will often refer to it as type IIB with O9/D9 sources. In addition to the O9-plane, one finds a web of O5-planes from the orbifold group $\Theta_m$ for $m=\alpha, \beta, \gamma, \alpha\beta, \alpha\gamma, \beta\gamma, \alpha\beta\gamma$. The internal transverse directions (–) and wrapped directions ($\times$) of the respective O-planes are summarized in the following table
\begin{align}\label{OplanesIIB}
\begin{pmatrix} 
&{\rm O}9:\quad & \times & \times & \times & \times & \times & \times & \times  \\
&{\rm O}5_{\alpha}:\quad & - & - & - & - & \times & \times & \times  \\
&{\rm O}5_{\beta} :\quad & - & - & \times & \times & - & - & \times  \\
&{\rm O}5_{\gamma}:\quad & - & \times & - & \times & - & \times & -  \\
&{\rm O}5_{\alpha\beta}:\quad & \times & \times & - & - & - & - & \times  \\ 
&{\rm O}5_{\beta\gamma} :\quad & \times & - & - & \times & \times & - & -  \\ 
&{\rm O}5_{\gamma\alpha} :\quad & \times & - & \times & - & - & \times & -  \\ 
&{\rm O}5_{\alpha\beta\gamma}:\quad & - & \times & \times & - & \times & - & - 
\end{pmatrix} \, . 
\end{align}

\section{\texorpdfstring{$\mathcal{N}=1$ supergravity theory in 3D}{}}

In this subsection, we start from the metric ansatz used to dimensionally reduce type II supergravity, introduce the unit-volume conventions for our setups, and then present the 3D effective theory constructed from $\mathcal{N}=1$ supergravity, which matches the theory obtained from the dimensional reduction.

Following the conventions of \cite{Farakos:2020phe} for dimensional reduction, we start from the 10-dimensional Einstein-frame action in \eqref{EinsteinFrameAction} and perform the reduction using the following metric ansatz
\be
\label{eq:metric}
ds^2_{10} =  e^{2\alpha \upsilon} ds^2_3 + e^{2\beta \upsilon}\widetilde{ds}^2_7\,,
\ee
where $\upsilon$ is the modulus describing the internal volume in the Einstein frame. Matching this to the definition in \eqref{volumeG2}, one finds
\begin{equation}
    \text{vol}(X)=e^{7\beta\upsilon}\,.
\end{equation}
The coefficients in the exponents are given by
\be\label{abrelation}
\alpha=-7\beta \,, \quad \beta = -\frac{1}{4 \sqrt{7}} \,, 
\ee
and this choice leads to the 3D Einstein frame with a canonical kinetic term for $\upsilon$.

In the metric \eqref{eq:metric}, the volume dependence of the internal space has been factored out from the internal-space metric $g_{mn}$, such that $g_{mn} = e^{2\beta \upsilon} \tilde{g}_{mn}$, where $\tilde{g}_{mn}$ is the metric of the unit-volume $G_2$ space.
For later convenience, we further extract the volume from the shape moduli and the radii by setting
\begin{equation}
    s^i=\text{vol}(X)^{3/7}\tilde{s}^i=e^{3\beta\upsilon}\tilde{s}^i\,,\quad\quad 
    r^i=\text{vol}(X)^{1/7}\tilde{r}^i=e^{\beta\upsilon}\tilde{r}^i\,.
\end{equation}
Since we factored out the volume of the internal space as a modulus, we find a constraint on the shape moduli from equation \eqref{volumeG2} above
\begin{equation}\label{unit}
\text{vol}(X) =e^{7\beta\upsilon} = e^{7\beta\upsilon} \lp\prod_{i=1}^7 \tilde{s}^i\rp^{\frac13} \qquad \Rightarrow \qquad
\prod_{i=1}^7 \tilde{s}^i = 1 \,\rightarrow \, \tilde{s}^7 = \prod_{a=1}^6 \frac{1}{\tilde{s}^a} \,.
\end{equation}
In all formulas we replace $\tilde{s}^7$ with the other six shape moduli using the above equation.


Starting from the 10D Einstein frame action in \eqref{EinsteinFrameAction} and considering the metric ansatz in \eqref{eq:metric} we obtain the 3D effective theory
\begin{equation}
e^{-1}\mathcal{L} = \frac{1}{2}R_3 - G_{IJ}\partial\varphi^I\partial\varphi^J - V(\varphi^I)\,,
\end{equation}
where capital indices $I, J = 1, \dots, 8$ label the eight scalar fields
\begin{equation}
\varphi^I = \upsilon, \phi, \tilde{s}^1, \dots, \tilde{s}^6 \,.
\end{equation}

The scalar potential in 3D supergravity can be written in terms of the superpotential $P(\varphi^I)$, which is a real function of the scalar fields, and the inverse field-space metric $G^{IJ}$ via
\begin{equation}\label{SUGRAscalarpotential}
V = G^{IJ} P_I P_J - 4P^2 \,,
\end{equation}
where $P_I = \partial_{\varphi^I}P$ denotes the derivatives of the superpotential with respect to the scalar fields. Considering the unit-volume constraint in \eqref{unit}, the moduli-space metric takes the following explicit form
\[
G_{IJ} =
\begin{pmatrix}
1/4 & 0   & 0 \\
0   & 1/4 & 0 \\
0   & 0   & \tilde{G}_{ab}
\end{pmatrix}
,\quad\quad
\tilde{G}_{ab} = \frac{1 + \delta_{ab}}{4 \tilde{s}^a \tilde{s}^b}, 
\quad\quad
a, b = 1, \dots, 6.
\tag{3.9}
\]
We refer to appendix \ref{app:DimensionalReduction} for more details about the dimensional reduction and note that we set $(2\pi)^2\alpha^{\prime}=1$.

\subsection{3D superpotential for IIA on G2-holonomy spaces}
We write down the superpotential of the type IIA theory constructed in \cite{Farakos:2020phe}. We do not delve into details here, but present it for completeness. The equations of motion and the properties for different sources and flux configurations can be found in \cite{Farakos:2020phe,Farakos:2023nms,Farakos:2023wps,Farakos:2025bwf}.
The general superpotential for type IIA compactifications on spaces with $G_2$-holonomy and with $F_0$, $F_4$ and $H_3$ fluxes is given by\footnote{We are using slightly different conventions from \cite{Farakos:2020phe}, see appendix \ref{app:DimensionalReduction} for details.} 
\begin{equation}
    P=
    \frac{1}{4 \text{vol}(X)^2} \int_X \lp e^{-\frac{\phi}{2}} \star\Phi\wedge H_3 + e^{\frac{\phi}{4}}\Phi \wedge F_4 + e^{\frac54 \phi}\star F_0 \rp \,.
\end{equation}
The analysis now becomes model dependent, as we consider the fluxes compatible with the orbifold of the compactification. In this construction, which serves as our starting point for building the type I AdS solution, the fluxes are expanded as follows
\begin{equation}\label{typeIIAfluxes}
H_3 = \sum_{i=1}^7 h^i \Phi_i \,,
\quad\quad
F_4 = \sum_{i=1}^7 f^i_4 \Psi_i \,,
\quad\quad
F_0 = m_0 \,.
\end{equation}
Since they are expanded in the closed and co-closed forms in equations \eqref{basis} and \eqref{basis2}, the fluxes are harmonic
\begin{equation}
    dH_3=0\,,\quad\quad dF_4=0\,, \quad\quad dF_0=0\,.
\end{equation}
Considering this flux ansatz, using $\text{vol}(X)=e^{7\beta \upsilon}$ and, after evaluating the integrals, the superpotential takes the following form
\begin{equation}
    P=\frac{1}{4} \lp e^{-\frac{\phi}{2}-10\beta\upsilon}\sum_{i=1}^7\frac{h^i}{\tilde{s}^i}
    +e^{\frac{\phi}{4}-11\beta\upsilon}\sum_{i=1}^7f^i_4\tilde{s}^i
    + m_0 \, e^{\frac54 \phi -7\beta\upsilon}\rp
    \,,\quad \tilde{s}^7=\frac{1}{\prod_{a=1}^6\tilde{s}^a}\,.
\end{equation}
Using the expression for the scalar potential in \eqref{SUGRAscalarpotential}, one finds that the resulting potential matches the scalar potential obtained from dimensional reduction in appendix~\ref{app:DimensionalReduction}.

\subsection{3D superpotential for IIB on co-calibrated G2-structure}

The superpotential describing the type IIB theory emerging after applying three T-dualities to the previous type IIA setup was constructed in \cite{Emelin:2021gzx} and has the following form 
\begin{equation}
    P=\frac{1}{4 \text{vol}(X)^2} \int_X \lp e^{-\frac{\phi}{2}} F_7 -e^{\frac{\phi}{2}} \star \hat{\Phi} \w F_3 +\frac12 \hat{\Phi}\wedge d\hat{\Phi} \rp \,.
\end{equation}
The fluxes that are invariant under the orbifold and extend along the internal space are
\begin{equation}\label{basisF7F3}
    F_7=-\mathcal{G}d\eta^{1}\wedge\dots\wedge d\eta^{7}\,,\quad\quad F_3=\sum_if^i_3\hat{\Phi}_i \,,
\end{equation}
where the signs of the fluxes $f^i$ and $\mathcal{G}$ are fixed by T-dualities.
Evaluating the integrals the superpotential takes the explicit form 
\begin{equation}\label{superpotentialTYPEI}
    P=\frac{1}{4} \lp -\mathcal{G} e^{-\frac{\phi}{2}-14\beta\upsilon }- e^{\frac{\phi}{2}-10\beta\upsilon}\sum_{i=1}^7\frac{f^i_3}{\tilde{s}^i} +\frac12 e^{-8\beta\upsilon}\sum_{i,j}^7\tilde{s}^i\mathcal{M}_{ij}\tilde{s}^j \rp
    \,,\quad
    \tilde{s}^7=\frac{1}{\prod_{a=1}^6\tilde{s}^a}\,.
\end{equation}
Using the expression for the scalar potential in equation \eqref{SUGRAscalarpotential}, one finds that the resulting potential matches the scalar potential obtained from dimensional reduction in appendix \ref{app:DimensionalReduction} in equations \eqref{fun1IIB}–\eqref{fun4IIB}.\footnote{There is the usual caveat that we do not get the source contributions from the O5/D5 directly but rather the flux terms that cancel them. So, the matching requires us to use the tadpole cancellation condition that we will discuss below around equation \eqref{eq:F3tadpole}.}  

\subsection{Kaluza-Klein states}

To estimate whether the 3D effective theory can decouple from the extra dimensions, we compare the vacuum expectation value to the Kaluza–Klein masses. Since we want the latter to be heavy and thus unobservable, the following ratio should be small
\begin{equation}\label{scaleseparationcondition}
    \frac{\langle V\rangle}{m_{\text{KK}}^2}\sim \frac{L_{\text{KK}}^2}{L_{\text{AdS}}^2}\ll 1 \,.
\end{equation}
Our KK mass estimate will be obtained by comparing the KK mass associated with each radius of the toroidal space to the vacuum energy. Although the KK spectrum may differ due to the non-trivial internal geometry (e.g., twisting), this method remains the most conservative. By considering each radius individually, we ensure that the theory exhibits scale separation.

To estimate the KK masses, we impose periodic identifications on the toroidal internal space $y^i\sim y^i+1$, and write down the line element in the following way
\begin{equation}
    ds_{10}^2=e^{2\alpha\upsilon}ds_{3}^2+e^{2\beta\upsilon}\sum_{i=1}^7\tilde{r}_i^2dy_{i}^2 \,,
\end{equation}
where the seven unit-volume radii $\tilde{r}_i$ are not equal for anisotropic internal spaces.
The ten-dimensional dilaton can be expanded as
\begin{equation}
\phi(x^\mu, y^i) = \sum_{i=1}^7\sum_{n_i=0}^{\infty} \phi_{n_i,i}(x^\mu) \cos(2\pi n_i y^i) \,,
\end{equation}
and this expression is substituted into the ten-dimensional Einstein-frame action \eqref{EinsteinFrameAction}. Following the steps outlined in detail in \cite{Farakos:2023wps}, we find that the masses squared of the first Kaluza–Klein modes $\phi_{1,i}(x^\mu)$ are
\begin{equation}
    m^2_{\text{KK},i}\equiv m^2_{\phi_{1,i}}=(2\pi)^2\frac{e^{-16\beta\upsilon}}{\tilde{r}^2_i} \,.
\end{equation}
The ratios of the relevant scales, which determine whether scale separation is achieved, are given by
\begin{equation}\label{scaleseparationconditionexplicit}
    \frac{\langle V\rangle}{m^2_{\text{KK},i}}\sim \tilde{r}_i^2e^{16\beta\upsilon}\langle V\rangle\equiv r_{S,i}^2e^{-\frac{\phi}{2}}e^{14\beta\upsilon}\langle V\rangle \,,
\end{equation}
where we defined the string frame radii 
\begin{equation}\label{stringradii}
    r_{S,i}=e^{\frac{\phi}{4}}e^{\beta\upsilon}\tilde{r}_i\,.
\end{equation}
Scale separation is achieved if the expression in equation \eqref{scaleseparationconditionexplicit} is very small even for the largest internal radius $r_i = e^{\beta \upsilon}\tilde{r}_i$. Higher order $\alpha^{\prime}$-corrections from the underlying string theory can be neglected, if the string frame radii are all large in string units, $r_{S,i} \gg 1, \forall i$.

\section{\texorpdfstring{Scale-separated AdS$_3$ vacua from massive type IIA}{}}

We review the type massive IIA flux compactifications on $G_2$-holonomy spaces introduced in \cite{Farakos:2025bwf}, which yield scale separation in the large-flux limit and integer conformal dimensions. We consider two examples: the first has some flat directions, while the second, featuring a more involved flux ansatz that threads all the cycles, gives masses to all scalar fields. The $H_3$ flux is the same in both examples; we only adjust how many cycles the $F_4$ flux threads and the manner in which it does so. This setup also involves O2- and O6-planes. The first example cancels the O2-plane charge locally by adding D2-branes, while the second example uses also the flux contribution $H_3 \w F_4$.


A distinctive feature of one of these models is that the O2-charge is canceled by D2-branes and the $F_4$ flux quanta are all unbounded. This places the configuration in the class of \textit{trivially unbounded fluxes}, as classified in \cite{Tringas:2025uyg}. More precisely, the $H_3 \wedge F_4$ contribution to the tadpole vanishes because the fluxes $H_3$ and $F_4$ always share a common leg. This method of canceling the tadpole appears to be the key feature for obtaining integer conformal dimensions, since in other approaches using flux tuning, as in \cite{Farakos:2020phe, Farakos:2023nms, Farakos:2023wps} (classified as \textit{non-trivially unbounded fluxes} in \cite{Tringas:2025uyg}), such integer conformal dimensions do not arise.

\subsection{Minimal construction in type IIA}

We start by introducing the flux configuration for the first construction, which takes the following form
\begin{align}
    F_0&=m_0 \,,\\
    F_4&=-N\left(\Psi_2+\Psi_5+\Psi_6+n\Psi_7\right) \,, \label{F4minimal}\\
    H_3&=h\left(\Phi_1+\Phi_3+\Phi_4\right) \,. \label{H3minimal}
\end{align}
Comparing this with the basis expansion in \eqref{typeIIAfluxes}, we obtain the identifications $f_4^2=f_4^5=f_4^6=-N$, $f_4^7=-nN$, and so on.
The parameter $n$, which, when it deviates from unity $n\neq 1$, introduces a slight anisotropy which has only a minor impact on the stabilization and does not affect at all the conformal dimensions of the dual theory.

For the flux configuration in equations \eqref{F4minimal}-\eqref{H3minimal}, the wedge product of the $H_3$ and $F_4$ flux vanishes due to equation \eqref{eq:PhiiPsij}, $\int\Phi_i\wedge\Psi_j=\delta_{ij}$. As a result, the flux contribution to the $F_6$ Bianchi identity
\begin{equation}\label{eq:tadpoleO2}
0 =dF_6 = H_3 \wedge F_4 + J_{O2} + J_{D2} 
\end{equation}
vanishes identically. This means that the O2-plane charge has to be canceled by an appropriate number of D2-branes so that $J_{O2} + J_{D2}=0$. In this way, the $F_4$ flux remains unconstrained while $H_3$ will be bounded by the $F_2$ Bianchi identity that is given by
\begin{equation}
0 =dF_2 = H_3 \, F_0 + \sum_i \lp J_{O6_i} + J_{D6_i} \rp\,.
\end{equation}
The charge of the $O6$-planes wrapping the cycles dual to the $H_3$ flux in equation \eqref{H3minimal} is canceled by the $H_3\,F_0$ contributions and enters the scalar potential as a net contribution. The remaining charges of orientifolds wrapping cycles that are not dual to the $H_3$ flux, are canceled locally by an appropriate number of D6-branes
\begin{align}
0 &= H_3 \, F_0 +\sum_i J_{O6_i} \,, \quad\quad i=\alpha, \alpha \beta, \alpha \gamma  \,,  \label{Bianchi1}\\
0 &= J_{O6_i} + J_{D6_i} \,, \hspace{1.77cm} i=\beta, \gamma, \beta \gamma, \alpha \beta \gamma \label{Bianchi2} \,.
\end{align}
The $H_3$ and $F_0$ flux quanta are then bounded by the number of O6-planes due to the equation above.

\subsection{Non-minimal construction in type IIA}

To give a mass to all scalar fields, the $F_4$ flux was generalized in \cite{Farakos:2025bwf}, so that it threads all seven four-cycles
\begin{equation}\label{F4nonminimal}
\begin{split}
    F_4=&-N\left(\Psi_2+\Psi_5+\Psi_6+n\Psi_7\right)\\
    &+(Q+G)\Psi_1+(Q+G)\Psi_3+(Q-2G)\Psi_4 \,.
\end{split}
\end{equation}
Comparing this with \eqref{F4minimal}, we see that the first line coincide. Thus, when the $N$ flux quanta dominates parametrically, then the properties of the previous construction are recovered. However, all fields acquire non-vanishing masses due to the more generic $F_4$ flux.

Let us now discuss the involved tadpole cancellation and its relation to the conformal dimensions of dual operators in the CFT.
The $H_3$ flux configuration considered here is the same as in \eqref{H3minimal}, so the O6-plane tadpole cancellation proceeds as before. However, since $F_4$ has additional components, the O2-plane tadpole cancellation becomes more involved.
Let us first denote the components by indices corresponding to their distinct flux quanta, e.g. $F_4^G$ for the $F_4$ component with flux quanta $G$, since each part of the $F_4$ flux plays a different role in the behavior of the model:
\begin{equation}
F_4 = F_4^N + F_4^Q + F_4^G \,.
\end{equation}
Here, the $N$ flux appeared in the previous model, the $Q$ flux helps cancel the O2-plane tadpole and the $G$ and $Q$ flux give masses to the previously massless fields. As before, the $F_4^N \wedge H_3$ term, which appears in the tadpole cancellation condition (cf. equation \eqref{eq:tadpoleO2}) vanishes trivially. The term $F_4^{G}\wedge H_3$, also vanishes automatically due to the flux choice
\begin{equation}
    F_4^G\wedge H_3=hG\,\Phi_1\wedge\Psi_1+hG\,\Phi_3\wedge\Psi_3-2hG\,\Phi_4\wedge\Psi_4=0 \,.
\end{equation}
Hence, the $F^N_4$ and $F_4^G$ fluxes remain trivially unbounded in the tadpole, as in \cite{Farakos:2020phe}.
The remaining components are those with flux quanta $Q$, and they cancel the O2-plane charge
\begin{equation}\label{tadpoleIIAexplicit}
    0=\int F_4^Q\wedge H_3+\int(\mu_{O2}+\mu_{D2})J_7\quad\rightarrow\quad 0=3Qh-(16-N_{D2}) \,.
\end{equation}
Since the orientifold number is $N_{O2}=2^7$, see \cite{Farakos:2020phe}, one can take $N_{D2}=1$ as in \cite{Farakos:2025bwf} and set the NSNS flux quantum $h=1$. These choices would fix $Q=5$ by the tadpole cancellation condition above. However, one can also make other choices and we leave $h$ and $Q$ unspecified in our expressions below but take $Q h>0$. This flux choice leads to a net O2-plane contribution together with a net O6-plane contribution in the potential, which is consistent with the anisotropy of the internal space.

For this example, one generically finds no integer conformal dimensions for the operators in the dual CFT. Only in the parametric limit $G\to \infty$, does one obtain the same integer conformal dimensions as in the minimal construction above, see \cite{Farakos:2020phe} and~\cite{Farakos:2023nms}.

\section{\texorpdfstring{Scale-separated AdS$_3$ vacua from type I theory}{}}
We will now T-dualize the above IIA setup three times to a type IIB flux compactification on a $G_2$ structure space with metric fluxes, $F_3$ and $F_7$ fluxes, as well as O5/O9-planes. Emerging from these T-dualities to type I string theory, the $F_3$ flux has two pieces,
\begin{equation}
  F_3 = F_3^{\,\text{closed}} + F_3^{\,\text{non-closed}} \,.    
\end{equation}
Interestingly, we will see that the closed part arises from dualizing the unbounded and closed $F_4$ flux, while the non-closed part arises from the Romans mass, which is bounded by the O6-plane tadpole cancellation condition. The $F_7$ flux arises from an $F_4$ flux quantum, which is also unbounded and closed.

As we will see, the charge of some of the O5-planes will be canceled by $dF_3$, via the  following Bianchi identity,
\begin{equation}
dF_{3,i}^{\,\text{non-closed}} = J_{O5,i}\,.
\end{equation}
The remaining O5-planes have their charge canceled by the presence of D5-branes so that $J_{O5,i}+J_{D5,i}=0$. This is of course analogous to the case discussed above in equations \eqref{Bianchi1}, \eqref{Bianchi2}.

\subsection{Three T-dualities: NSNS sector}
The constructions we dualize share the same NSNS sector, so the metric fluxes generated from the Kalb–Ramond flux are identical. We present them here because they will be the same for both models.

To see how the $H_3$ flux transforms to metric flux under T-duality we consider the basis in equation \eqref{basis} and the flux ansatz in equation \eqref{H3minimal}.
It is convenient to express the flux by its definition
\begin{equation}
    H_3 = \frac{1}{3!}H_{ijk}\eta^{ijk} \,,
\end{equation}
and the according to the basis in \eqref{basis} flux components in our configuration are
\begin{equation}\label{Hdual1}
    H_{127}= h\,,\quad\quad H_{567}= -h \,,\quad\quad H_{136}= h \,.
\end{equation}
Using the rules from \cite{Shelton:2005cf}, we perform the three T-dualities along the directions $y_2, y_3$, and $y_5$.
For a general flux $H_{ijk}$, if we want to dualize along the $i$ direction we get
\begin{equation}
    H_{ijk}\leftrightarrow \tau^i_{jk} \,,
\end{equation}
while along $j$ direction we get
\begin{equation}
    H_{ijk}\leftrightarrow -H_{jik} \leftrightarrow -\tau^j_{ik} \,.
\end{equation}
Thus, the three NSNS fluxes in equation\eqref{Hdual1} give the following metric fluxes
\begin{equation}
    H_{127}\leftrightarrow -\tau^2_{17}\equiv h \,,\quad\quad
    H_{567}\leftrightarrow \tau^5_{67}\equiv-h \,,\quad\quad
    H_{136}\leftrightarrow -\tau^3_{16}\equiv h\,.
\end{equation}
The metric flux matrix in equation \eqref{mijmatrix}, takes the form
\be\label{matrixexample1}
{\cal M}_{ij} = \begin{pmatrix}
  0 & 0 & 0 & 0 & 0 & 0 & 0 \\ 
  0 & 0 & 0 & 0 & -\tau^{3}_{16} & 0 & 0 \\
  0 & 0 & 0 & 0 & 0 & 0 & 0 \\
  0 & 0 & 0 & 0 & 0 & 0 & 0 \\
  0 & -\tau^{3}_{16} & 0 & 0 & 0 & -\tau^{5}_{67} & -\tau^{2}_{17} \\
  0 & 0 & 0 & 0 & -\tau^{5}_{67} & 0 & 0 \\
  0 & 0 & 0 & 0 & -\tau^{2}_{17} & 0 & 0
\end{pmatrix} \,, 
\ee
with the non-zero entries given by $\mathcal{M}_{52}$, $\mathcal{M}_{56}$, $\mathcal{M}_{57}$, and their symmetric counterparts.
We see that the non-zero metric fluxes are all equal, which is expected since the NSNS flux in the type IIA theory also had equal components
\begin{equation}\label{metricfluxes}
\mathcal{M}_{52}=\mathcal{M}_{56}=\mathcal{M}_{57}=\mathcal{M}_{25}=\mathcal{M}_{65}=\mathcal{M}_{75}= h \,.
\end{equation}
It is worth noting that the group structure obtained by T-dualizing the type IIA setup of \cite{Farakos:2025bwf} corresponds to the nilmanifold $\mathfrak{n}_{10}$ in the classification of \cite{VanHemelryck:2025qok} (Table 3 there) \footnote{Importantly, that work did not find a classical solution with scale separation on $\mathfrak{n}_{10}$ because the flux configuration and the way tadpoles are canceled here differ from those used there. Although that paper does not present the solution for this nilmanifold explicitly (only the conclusion appears in Table 3), our discussion with the author clarified that some of our unbounded fluxes map to fixed fluxes in their setup, so it does not yield our solution. The background there can also be interpreted as type I after removing orbifold shifts, thereby allowing the inclusion of the O9/D9 system.}.

\subsection{Dualizing a minimal isotropic construction}

In this subsection we present the RR fluxes and sources obtained from the three T-dualities. We identify which fluxes are constrained by the tadpole cancellation condition and which are unconstrained. 
 
To construct our first example, we act on the type IIA $F_4$ flux in \eqref{F4minimal} from the left with the three T-dualities, which can be represented by the operator 
\begin{equation}\label{Tdualities1}
    T_{y^5}T_{y^3}T_{y^2} \,,
\end{equation}
see Appendix \ref{Dualfieldsexplicit} for the detailed action on the fluxes.
The $F_4$ flux components map to $F_3$ and $F_7$ fluxes, while the Romans mass becomes an $F_3$ flux. Thus, the fluxes on the type I side, expanded on the relevant basis, take the form (see Appendix \ref{Dualfieldsexplicit} for a detailed derivation and conventions)
\begin{equation}\label{F3minimal}
    F_3=nN\hat{\Phi}_1+N\hat{\Phi}_3+N\hat{\Phi}_4-\tilde{m}_0\hat{\Phi}_5 \,,\quad\quad
    F_7=-\mathcal{G}d\eta^{1}\wedge\dots\wedge d\eta^{7} \,.
\end{equation}
Above, we have replaced $m_0$ with $\tilde{m}_0=-m_0$ for later convenience. The original $F_4$ flux in equation \eqref{F4minimal} was also generalized by changing the prefactor of $\psi_5$ from $-N$ to $-\mathcal{G}$. The reason is that there are interesting solutions, if we allow the $F_7$ flux to scale differently. The reason for our particular sign choices is that later all relative signs between the flux quanta will be fixed in the supersymmetric vacuum and we can choose all flux quanta to be positive
\begin{equation}
N,\; n,\; \tilde{m}_0,\; \mathcal{G}, \; h \;>\; 0 \,.
\end{equation}

The currents of the O6-planes and D6-branes transform to O5-planes and D5-branes, generally. They enter the Bianchi identity in the following way
\begin{equation}\label{eq:F3tadpole}
    dF_3=\sum_i \lp \mu_{O5}J_{4,i}+\mu_{D5}J_{4,i} \rp\,,
\end{equation}
where $i$ runs over all indices produced by the orbifold images, see Table~\ref{OplanesIIB}.
More specifically, the O6-planes whose charges are not canceled by D6-branes but rather by fluxes give rise, on the type I side, to the following O5-plane currents (see Table~\ref{IIAIsources}),
\begin{align}\label{O5currents}
    J_{4,\beta}=\hat{\Psi}_{2} \,,
    \quad\quad
    J_{4,\gamma}=\hat{\Psi}_{7} \,,
    \quad\quad
    J_{4,\beta\gamma}=\hat{\Psi}_{6} \,.
\end{align}
These contribute to the potential, providing the crucial negative terms responsible for moduli stabilization and scale separation, see \cite{Gautason:2015tig,Tringas:2025uyg}.
The O6$_{\alpha\beta\gamma}$/D6$_{\alpha\beta\gamma}$ system transforms to O9/D9, see Table~\ref{IIAIsources}, and this is how wet get to type I theory.

Now we write down the explicit Bianchi identities. To calculate the non-trivial Bianchi identities we decompose the exterior derivative of $F_3$ into components
\begin{equation}
    dF_3=\sum_idF_3^{(i)}\equiv \sum_if^i_3 d\hat{\Phi}_i=nNd\hat{\Phi}_1+Nd\hat{\Phi}_3+Nd\hat{\Phi}_4-\tilde{m}_0d\hat{\Phi}_5\,,\\
\end{equation}
where $d\Phi_i$ is not necessarily zero; its value depends on the non-vanishing metric fluxes in \eqref{mijmatrix}. 
To explicitly compute the non-vanishing components, we use the relation \eqref{dPhi} together with the metric flux matrix in \eqref{matrixexample1}. We observe that the metric fluxes appearing in the exterior derivative of the relevant fluxes $\mathcal{M}_{1j}$, $\mathcal{M}_{3j}$, and $\mathcal{M}_{4j}$ are zero, and consequently
\begin{align}
    dF^{(1)}_3=dF^{(3)}_3=dF^{(4)}_3= 0 \,,
\end{align}
which implies that the components $f_3^1=nN$, $f_3^3=N$, and $f_3^4=N$ of $F_3$ do not contribute to the tadpole cancellation and are therefore not bounded.
On the other hand, the non-zero metric fluxes contribute as follows
\begin{equation}\label{eq:dF35}
    dF^{(5)}_3=f^5_3\sum_j\mathcal{M}_{5j}\hat{\Psi}_j
    =f^5_3\mathcal{M}_{52}\hat{\Psi}_2+f^5_3\mathcal{M}_{56}\hat{\Psi}_6+f^5_3\mathcal{M}_{57}\hat{\Psi}_7\,.
\end{equation}
These components enter the Bianchi identity in precisely the way required to cancel the O5-plane currents above in equation \eqref{O5currents},
\begin{equation}\label{eq:F35explicit}
    f^5_3\mathcal{M}_{52}\hat{\Psi}_2=\mu_{O5}J_{4,\beta} \,,\quad
    f^5_3\mathcal{M}_{56}\hat{\Psi}_6=\mu_{O5}J_{4,\beta\gamma} \,,\quad
    f^5_3\mathcal{M}_{57}\hat{\Psi}_7=\mu_{O5}J_{4,\gamma} \,.
\end{equation}
Since the O5-plane charge does not scale, the combination
\begin{equation}
    f^i_3\mathcal{M}_{ij}=\mu_{O5}\sim N^0 \,,
\end{equation}
must not scale either. To satisfy flux quantization consistently, fluxes must either scale positively (becoming large) or not scale at all. In our case, the fluxes involved in the cancellation must therefore obey
\begin{equation}
    f^5_3\sim N^0\,,\quad \mathcal{M}_{ij}\sim N^0 \,. 
\end{equation}
It is interesting, though expected, that the four unbounded $F_4$ flux components in type IIA become three $F_3$ fluxes and one $F_7$ flux, all of which remain unbounded, while the $H_3$ flux bounded by the O6-plane tadpole becomes a metric flux bounded by the O5-plane tadpole.

\subsection{Moduli stabilization and parametric behavior}\label{minimalanisotropic}

To proceed, we take the fluxes obtained from the three T-dualities, the metric fluxes in equation \eqref{metricfluxes} and the RR fluxes in equation \eqref{F3minimal}, and substitute them into the general type IIB superpotential in equation \eqref{superpotentialTYPEI}. Then we extremize it
\begin{equation}
    \partial_{\upsilon}P=0\,,\quad\quad
    \partial_{\phi}P=0\,,\quad\quad
    \partial_{\tilde{s}^a}P=0\,,\quad\text{for}\quad a=1\,,\dots\,,6\,.
\end{equation}
By solving this system of eight equations, we obtain the following relations for the moduli at the vacuum
\begin{equation}\label{modulifluxes1}
    \tilde{s}^1=n\tilde{s}^3\,,
    \quad\quad
    \tilde{s}^3=\tilde{s}^4\,,
    \quad\quad
    \tilde{s}^2=\tilde{s}^6\,,
    \quad\quad
    \tilde{s}^5=\frac{1}{n(\tilde{s}^2\tilde{s}^3)^3}\,.
\end{equation}
From the explicit expression of the third cycle (analogously for the others), 
\begin{equation}\label{eq:stabmodel1}
    \tilde{s}^3=\left(\frac{2N}{\tilde{m}_0n(\tilde{s}^2)^3}\right)^{1/4}\,,\quad
    e^{\beta\upsilon}=\frac{\mathcal{G}^{\frac18}}{h^{\frac14}}\lp \frac{2^5 N^9 n^3 \tilde{s}^2}{\tilde{m}_0^5}\rp^{\frac{1}{32}}\,,\quad
    e^{\phi}=\frac{h \mathcal{G}^{\frac12}}{2^{\frac38}} \left(\frac{\tilde{m}_0^3}{N^{15} n^5 (\tilde{s}^2)^7}\right)^{1/8}\,,
\end{equation}
we see that the moduli cannot be written solely in terms of fluxes.
This implies that one modulus can take arbitrary values and remains a flat direction. We will see this in the mass spectrum, where this unfixed $\tilde{s}^2$ scalar field implies the existence of at least one massless scalar field.

The flux components in \eqref{F3minimal} yield a model with desirable properties, despite the presence of an unstabilized modulus. Following \cite{Farakos:2025bwf}, we temporarily set fields equal $\tilde{s}^2 = \tilde{s}^3 = \tilde{s}^4 = \tilde{s}^6$, while in the next section we stabilize them consistently using additional fluxes. 

We solve the first equation in equation \eqref{eq:stabmodel1} above for $\tilde{s}^3=\tilde{s}^2$ to find
\begin{align}\label{modulifluxes2}
    \tilde{s}^2=\left(\frac{2N}{\tilde{m}_0n}\right)^{1/7}
    \,,\quad\quad
    e^{\beta\upsilon}
    =\left(\frac{2^9n^5}{\tilde{m}_0^9h^{14}}\right)^{1/56}N^{2/7}\mathcal{G}^{1/8}
    \,,\quad\quad
    e^{\phi}
    =\sqrt{\frac{\tilde{m}_0h^2}{2n}}\frac{\mathcal{G}^{1/2}}{N^2}\,,
\end{align}
and we remind the reader that the fluxes $N$ and $\mathcal{G}$ are the only ones that are unbounded and can be taken parametrically large. 
Now, evaluating the superpotential at the vacuum, one can calculate the vacuum expectation value to be
\begin{align}\label{vacuum1}
    \langle V\rangle
    =-4P^2\big\vert_{\text{SUSY}}  
    =-\frac{\mathcal{G}^2}{4}e^{-28\beta\upsilon-\phi}
    =-\frac{1}{2^{6}}\frac{\tilde{m}_0^{4}h^6}{n^2}\frac{1}{N^{6}\mathcal{G}^2} \,.
\end{align}
Before inserting the explicit solutions, the second relation describes a general AdS vacuum and agrees with \cite{Emelin:2021gzx}. The key difference is that, in our setup, the $F_7$ flux quantum $\mathcal{G}$ is unconstrained by the equations of motion, a consequence of the group-manifold structure. This strongly affects the string coupling and the cycle volumes, and provides additional freedom for new solutions, as we will see.

\subsection{Families of solutions}
Since we have two unbounded fluxes, $\mathcal{G}$ from $F_7$ and $N$ from $F_3$, and all moduli and vacuum expectation values are expressed in terms of them, we aim to determine the flux conditions that yield weak or strong coupling, large or small volume, and scale separation. 

We first examine the string coupling in equation \eqref{modulifluxes2} and observe that it is controlled by these two fluxes in an inverse manner, which parametrically implies that
\begin{equation}
    e^{\phi}\sim \mathcal{G}^{1/2}N^{-2} \,.
\end{equation}
Thus, when the flux $\mathcal{G}$ dominates as described below, the solution becomes strongly coupled at large flux values, whereas if $N$ dominates, the theory remains weakly coupled
\begin{equation}\label{weakstringcoupling}
    \text{weak coupling}\quad:\quad \mathcal{G}\ll N^4\,,
    \quad\quad
    \text{strong coupling}\quad:\quad\mathcal{G}\gg N^4\,.
\end{equation}
To probe the parametric behavior of the internal lengths, we analyze the string-frame volumes of the three-cycles and obtain 
\begin{align}
    s^1=\frac{2^{1/4}(n\mathcal{G})^{3/4}}{\tilde{m}_0^{1/4}N^{1/2}}\,,\quad\quad s^5=\frac{\tilde{m}_0}{2nN}s^1\,,\quad\quad s^2=s^3=s^4=s^6=s^7\equiv\frac{1}{n}s^1 \,.
\end{align}
Since $n$ is a deformation parameter, it is straightforward to see that, parametrically, all cycles become equal except $s^5$.
Ignoring the small bounded flux prefactors, this leads to two parametric regimes:
\begin{equation}\label{cycles2}
    \text{All cycles large:}\quad N\ll\mathcal{G}^{1/2}\,,
    \quad\quad
    \text{all large but $s^5$ small:}\quad\mathcal{G}^{1/2}\ll N\ll \mathcal{G}^{3/2}\,.
\end{equation}
The first regime yields a new class of classical solutions, whereas the second reproduces the T-dual of \cite{Farakos:2025bwf}. We discuss both cases in the brief classification below.

Let us now compute the behavior of the string-frame radii, which we need to estimate scale separation. Explicitly, we find
\begin{align}
    r_{S,\{1,7\}}=n^{1/2}r_{S,\{4,6\}}=n^{1/4}\left(\frac{2\mathcal{G}}{\tilde{m}_0}\right)^{1/4}, 
    \quad\hspace{0.25cm}
    r_{S,2}=n^{1/2}r_{S,\{3,5\}}=n^{1/4}\left(\frac{\tilde{m}_0\mathcal{G}}{2N^2}\right)^{1/4},  
\end{align}
and parametrically, since $n$ is as a deformation parameter, the internal sizes are set by two distinct scales, controlled by different flux combinations.

We can now calculate whether we can get scale separation and since our internal space has two distinct scales, reflecting the different behavior of the radii, we compare them to the vacuum expectation value to obtain
\begin{equation}\label{scaleseparation1}
\begin{split}
    \frac{\langle V\rangle}{m_{\text{KK},\{1,7\}}^2}&
    =n\frac{\langle V\rangle}{m_{\text{KK},\{4,6\}}^2}\sim \tilde{m}_0h^2 \frac{1}{N} \,, \\
    \frac{\langle V\rangle}{m_{\text{KK},2}^2}&
    =n\frac{\langle V\rangle}{m_{\text{KK},\{3,5\}}^2}
    \sim \tilde{m}_0^2h^2\frac{1}{N^2} \,.
\end{split}
\end{equation}
It is obvious that for parametrically large $N$ we always get scale separation\footnote{It is important to note that we examine the scaling behavior of each radius relative to the characteristic scale of the vacuum, which guarantees scale separation. This is a very strict condition that we impose even though there are no dynamical one-cycles in this setup, which would be characterized by a single radius. However, since we do not perform the exact KK analysis for the twisted tori, we use this strict criterion to ensure that we indeed achieve scale separation.
This is also a key difference compared to the solutions in \cite{Arboleya:2024vnp,VanHemelryck:2025qok}, where certain lengths become parametrically small, potentially causing issues in solvmanifold constructions but not in nilmanifolds, as argued in \cite{VanHemelryck:2025qok}.}.
The only fluxes that control scale separation are three of the four unbounded components $N$ of $F_3$, which can be taken to infinity to achieve parametric scale separation, satisfying equation \eqref{scaleseparationcondition}. The 7-form flux filling the entire internal space drops out, and with only this flux present, analogous to well-known compactifications such as Freund–Rubin or $\text{AdS}_5 \times S^5$ \cite{Freund:1980xh, Schwarz:1983qr}, scale separation could not be achieved. 

We now present the conditions of the fluxes that lead to different solutions of the theory and the parametric behavior of each solution. We find the following three families of solutions
\begin{itemize}
    \item Parametric classical solution with scale separation:
    \begin{equation}\label{classicalsolution}
        1\ll N^2\,\ll\,\mathcal{G}\,\ll\,N^4 \,.
    \end{equation}
    This branch yields a new type I solution in which the string coupling is parametrically small, all three-cycles are parametrically large, and scale separation is achieved. It T-dualizes to a type IIA solution with parametrically shrinking cycles. The existence follows from allowing for an additional scaling freedom in the $F_7$ flux, (T-dual to $F_4$). The corresponding $\mathcal{G}$ flux quanta which controls the string coupling $g_s$ via equation \eqref{weakstringcoupling} and the cycle sizes together with $F_3$ via equation \eqref{cycles2}.  Scale separation is governed by only the $F_3$ fluxes via equation~\eqref{scaleseparation1}.
    \item Parametric weak coupling, dual $s^5$ cycle shrinking, with scale separation:
    \begin{equation}\label{dualsolution}
        1\ll\,\mathcal{G}^{1/2}\,\ll\,N\,\ll\,\mathcal{G}^{3/2} \,.
    \end{equation}
    This solution is the T-dual of the classical type IIA background of \cite{Farakos:2025bwf}, albeit slightly more general due to the additional freedom in the flux sector.
    
    \item Parametric strongly coupled solution, all cycles large and scale separation:
    \begin{equation}\label{stronglycoupledsolution}
       1 \ll N^4\,\ll\,\mathcal{G} \,.
    \end{equation}
    This is a new type I solution in which the $F_7$ flux dominates over $F_3$, as described in equation \eqref{weakstringcoupling}. It corresponds to a weakly coupled heterotic solution, and its generalization with all moduli stabilized by fluxes, will appear soon in \cite{Het}.
\end{itemize}
We note once more that, in the solutions above, an unstabilized modulus was fixed by hand to a flux dependent value. In the next section we stabilize this modulus by introducing additional fluxes, thereby recovering a more general family of solutions that preserves the desired properties presented here.

\subsubsection{Masses and dimensions of dual operators}

To calculate the conformal dimensions of the dual CFT for the above construction with the flux configuration in equation \eqref{F3minimal}, we first compute the derivatives of the scalar potential in equation \eqref{SUGRAscalarpotential} to obtain the Hessian $V_{IJ}$. Using that the AdS length for our AdS vacuum is given by $L^2= \langle V \rangle^{-1}$ we find 
\begin{equation}\label{eigen1}
    m^2L^2=\frac12 \text{Eigen}\left[\langle G_{IJ}\rangle^{-1}\frac{\langle V_{IJ}\rangle}{\vert\langle V\rangle\vert}\right]=\left\{48,8,8,8,8,0,0,0\right\} \,.
\end{equation}
The conformal dimensions for the corresponding operators in the dual CFT are integers
\begin{equation}
    \Delta=1+\sqrt{1+m^2L^2}=\left\{8,4,4,4,4,2,2,2\right\} \,.
\end{equation}
These results match exactly those of the first isotropic example in type IIA \cite{Farakos:2025bwf}, which is expected since the two theories are T-dual. The zero eigenvalues found in equation \eqref{eigen1} correspond to massless scalars, and therefore the solution is not truly scale-separated, since there is a moduli space. However, as in \cite{Farakos:2025bwf}, this can be remedied by considering a more involved flux ansatz for $F_4$, which in our solution corresponds to a more complicated $F_3$ flux that we will discuss next.

\subsection{T-dual vacua from anisotropic type IIA and new type I solutions}\label{maximalanisotropic}

Performing the three T-dualities along the directions \eqref{Tdualities1}, the non-minimal $F_4$ flux ansatz in \eqref{F4nonminimal} and the Romans mass yield the following dual $F_3$ flux in the type I theory
\begin{equation}\label{F3anisotropic2}
\begin{split}
    F_3=&nN\hat{\Phi}_1+N\hat{\Phi}_3+N\hat{\Phi}_4-\tilde{m}_0\hat{\Phi}_5\\
    &-(Q+G)\hat{\Phi}_6-(Q+G)\hat{\Phi}_7-(Q-2G)\hat{\Phi}_2 \,,
\end{split}
\end{equation}
where all the flux components are positive, see Appendix \ref{Dualfieldsexplicit} for details on the duality.
The flux $N$, which appear in the ansatz \eqref{F3anisotropic2}, and $\mathcal{G}$ in $F_7=-\mathcal{G}d\eta^{1}\wedge\dots\wedge d\eta^{7} $ appeared already in the flux configuration in equation \eqref{F3minimal} above. They both remain unbounded but we now have two new $F_3$ flux quanta $Q,G$ and as we will see below $G$ will actually also be unbounded in our solution.
The metric fluxes are still given by the matrix in equation \eqref{matrixexample1}, since the NSNS sector of the type IIA theory remains unchanged for all out examples.

Let us discuss the Bianchi identities related to the O5-planes. In addition to $dF_3^{(5)}$ in equations \eqref{eq:dF35} and \eqref{eq:F35explicit} above, we have now more terms appearing in $dF_3$. Using the relation in equation \eqref{dPhi} along with the metric flux matrix in equations \eqref{matrixexample1} and \eqref{metricfluxes} we find 
\begin{equation}
\begin{split}
    dF_3=& dF_3^{(5)}+ dF^{(2)}_3+dF^{(6)}_3+dF^{(7)}_3\\
    =& dF_3^{(5)}+ h\big((Q-2G)+(Q+G)+(Q+G)\big)\hat{\Psi}_5 \\
    =& dF_3^{(5)}+ 3h\,Q\hat{\Psi}_5\,.
\end{split}
\end{equation}
We notice that the $Q$ flux quantum in the $F_3$ flux is thus bounded by this tadpole. However, we see that the $G$ flux cancels in the same way as in type IIA and therefore remains unbounded. 
So we have identified the bounded fluxes as the metric fluxes $h$ together with the $\tilde{m}_0$ and $Q$ components of $F_3$, dual to the $H_3$, Romans mass, and $Q$ components of $F_4$ in the type IIA setup.

Since five distinct $F_3$ flux quanta appear in equation \eqref{F3anisotropic2}, the anisotropy is larger than before, and consequently the radii will be more anisotropic compared to the previous minimal example. Plugging the flux ansatz into the superpotential and solving the equations of motion, we obtain the following relations among the moduli
\begin{equation}\label{cycle2346ANIS2}
\begin{split}
    \tilde{s}^1&=n\tilde{s}^3
    \,,\quad\quad
    \tilde{s}^2=\frac{4G^2-Q^2}{2NQ}\tilde{s}^3
    \,,\quad\quad
    \tilde{s}^{3}=\tilde{s}^{4}\,,\\ 
    \tilde{s}^5&=\frac{\tilde{m}_0}{2N}\tilde{s}^3
    \,,\hspace{0.42cm}
    \tilde{s}^6=\frac{(G+Q)(2G+Q)}{NQ}\tilde{s}^3=\tilde{s}^7 \,.
\end{split}
\end{equation}
Now, requiring the moduli to be positive and choosing the branch with $N>0$, we obtain the following conditions on the $F_4$ flux component $G$ and $Q$:
\begin{equation}
    G<-Q<0 \,,\quad\text{or}\quad G>\frac{Q}{2}>0 \,.
\end{equation}
To avoid signs and cumbersome notation below, we choose the second branch with $Q,G>0$. We explicitly express one of the moduli entirely in terms of fluxes, so that its value can then be substituted into the previous relations to determine all moduli in terms of fluxes
\begin{equation}\label{s3UNIT}
    \tilde{s}^3=\frac{2^{2/7}Q^{3/7}N^{4/7}}{\left(n \tilde{m}_0(2G-Q)(G + Q)^2 (2 G + Q)^3\right)^{1/7}} \,.
\end{equation}
Next, we present the dilaton and the extracted Einstein-volume of the radii, which are stabilized by fluxes, as shown below
\begin{align}
    e^{\phi}&=\frac{h}{2(G+Q)}\sqrt{\frac{\tilde{m}_0\mathcal{G}Q(2G+Q)}{nN^3(2G-Q)}} \,,\label{dilatonfull}\\
    e^{\beta\upsilon}&=\frac{2^{9/28}\mathcal{G}^{1/8}(G+Q)^{5/28}(nN^{3}(2G-Q))^{5/56}}{h^{1/4}\big(\tilde{m}_0^9Q(2G+Q)^{13}\big)^{1/56}}\,,
\end{align}
and we see that in this construction all moduli are stabilized by fluxes.
Now, using the radii expressed in terms of three-cycles $s^i$, we calculate the string-frame radii through the Weyl rescaling in \eqref{stringradii}, obtaining
\begin{align}
    r_{S,1}=\frac{\sqrt{2}}{\tilde{m}_0^{1/4}}\left(\frac{nN\mathcal{G}Q}{4G^2-Q^2}\right)^{1/4} \,,
\end{align}
and express the rest radii in terms of it
\begin{align}
    r_{S,2}&=\sqrt{\frac{\tilde{m}_0}{2N}}r_{S,1} \,,
    \quad\quad
    r_{S,3}=\sqrt{\frac{\tilde{m}_0}{4nN}}\sqrt{\frac{2G-Q}{G+Q}}r_{S,1} \,, \\
    r_{S,4}&=\frac{(2G+Q)\sqrt{(2G-Q)(G+Q)}}{\sqrt{2 n}\, NQ}r_{S,1}\,,
    \quad\quad
    r_{S,5}=\sqrt{\frac{\tilde{m}_0}{2nN}}r_{S,1} \,, \\
    r_{S,6}&=\frac{1}{\sqrt{n}}r_{S,1}\,,
    \quad\quad
    r_{S,7}=\sqrt{\frac{2G-Q}{2(G+Q)}} r_{S,1} \,.
\end{align}
The ratios of the string-frame three-cycle moduli coincide with the unit-volume ratios in \eqref{cycle2346ANIS2}. Their absolute values, however, differ by an overall flux-dependent rescaling. Accordingly, we fix one cycle explicitly and obtain the rest from the ratio relations. The string-frame value of $s^3$ is
\begin{equation}\label{s3string}
    s^3= \left(\frac{4 \mathcal{G}^3 Q^3 N}{n\tilde{m}_0 (2G-Q) (G+Q)^2 (2G+Q)^3}\right)^{1/4}\,,
\end{equation}
and the remaining $s^i$ follow directly from \eqref{cycle2346ANIS2}, e.g. the first cycle modulus will be $s^1=ns^3$. 
We write down the vacuum expectation value in terms of the fluxes and its parametric behavior
\begin{equation}
    \langle V\rangle=-4P^2=-\frac{h^6\tilde{m}_0^4(Q+2G)^6}{2^{10}n^2(2G-Q)^2(Q+G)^4}\frac{1}{\mathcal{G}^2N^6}
    \,\sim\,
    -\frac{1}{\mathcal{G}^2N^6}\,.
\end{equation}
We observe that, at the parametric limit, the dependence on the unbounded $G$ flux drops out. The result depends only on the fluxes of the previous setup, and the vacuum expectation value has exactly the same parametric form as in equation \eqref{vacuum1}.

\subsection{Families of solutions}
In this subsection we determine the flux conditions that lead to weak or strong coupling, large or small volume, and scale separation. In this non-minimal setup, besides the two unbounded fluxes $\mathcal{G}$ (from $F_7$) and $N$ (from $F_3$), there is an additional unbounded $F_3$ component $G$.

We examine the string coupling in \eqref{dilatonfull} and observe that there is more flux which contributes to the controlled regimes of the coupling, which parametrically implies that
\begin{equation}\label{dilatonscaling}
     g_s\equiv e^{\phi}\sim\frac{1}{G}\sqrt{\frac{\mathcal{G}}{N^3}} \,, 
\end{equation}
and the weakly and strongly coupled regimes are determined parametrically by the following conditions
\begin{equation}\label{weakstringcoupling2}
    \text{weak coupling}\quad:\quad \mathcal{G}\ll N^3G^2\,,
    \quad\quad
    \text{strong coupling}\quad:\quad\mathcal{G}\gg N^3G^2\,.
\end{equation}
To determine whether the cycles become parametrically large or small, we use the string-frame versions of the ratios in \eqref{cycle2346ANIS2} together with the explicit expression \eqref{s3string}. Taking the parametric limit and dropping bounded fluxes and numerical prefactors, we obtain two regimes: 
\begin{itemize}\label{cyclesnonminimal}
    \item All cycles large: $\mathcal{G} \gg G^{2} N$ \,,
    \item $s^5$ parametrically small, all other cycles large: $\mathcal{G} \ll G^2N$, $N^3 \mathcal{G}^{-3}\ll G^2\ll\mathcal{G}N^{\frac13}$.
\end{itemize}
The second limit with parametrically small $s^5$ is T-dual to the type IIA solution in \cite{Farakos:2025bwf}.
Next, we present the ratios \eqref{scaleseparationconditionexplicit} of the vacuum expectation value for each of the radii. For the first radius we obtain
\begin{align}
    \frac{\langle V\rangle}{m_{\text{KK},1}^2} \sim \frac{\tilde{m}_0h^2(2G+Q)^2}{N(2G-Q)(G+Q)}\,,
\end{align}
and express the remaining ratios in terms of it
\begin{equation}
\begin{split}
    \frac{\langle V\rangle}{m_{\text{KK},1}^2}
    &=n\frac{\langle V\rangle}{m_{\text{KK},6}^2}
    =\frac{2(G+Q)}{2G-Q}\frac{\langle V\rangle}{m_{\text{KK},7}^2} \,,\\
    \frac{\langle V\rangle}{m_{\text{KK},1}^2}
    &=\frac{2N}{\tilde{m}_0}\frac{\langle V\rangle}{m_{\text{KK},2}^2}
    =\frac{4nN(G+Q)}{\tilde{m}_0(2G-Q)}\frac{\langle V\rangle}{m_{\text{KK},3}^2}
    =\frac{2nN}{\tilde{m}_0}\frac{\langle V\rangle}{m_{\text{KK},5}^2}
    \,, \\
    \frac{\langle V\rangle}{m_{\text{KK},1}^2}
    &=\frac{2nN^2Q^2}{(2G-Q)(G+Q)(2G+Q)^2}\frac{\langle V\rangle}{m_{\text{KK},4}^2} \,.
\end{split}
\end{equation}
To avoid cumbersome expressions when investigating different parametric regimes, we ignore overall numerical coefficients and bounded fluxes and keep the parametric dominant ones to find
\begin{equation}\label{scaleseparation2}
    \frac{\langle V\rangle}{m_{\text{KK},\{1,6,7\}}^2}\sim \frac{1}{N} \,,\quad\quad
    \frac{\langle V\rangle}{m_{\text{KK},\{2,3,5\}}^2}\sim \frac{1}{N^2} \,,\quad\quad
    \frac{\langle V\rangle}{m_{\text{KK},4}^2}\sim \frac{G^4}{N^3} \,.
\end{equation}
We now present the conditions of the fluxes that lead to different solutions of the theory and the parametric behavior of each solution. We find the following three families of solutions:
\begin{itemize}
    \item Parametric classical solution with scale separation: 
    \begin{equation}
          G^{\frac{10}{3}}\ll G^2N\ll\mathcal{G}\ll G^2N^3\,.
    \end{equation}
    This branch yields a new type I solution in which the string coupling is parametrically small, all three-cycles are parametrically large, and scale separation is achieved. It T-dualizes to a type IIA solution with parametrically shrinking cycles. The construction follows from allowing additional freedom in the $F_7$ flux, (T-dual to $F_4$) which controls $g_s$ \eqref{weakstringcoupling2} and the cycle sizes together with $F_3$ \eqref{cyclesnonminimal}, while scale separation is governed by only the $F_3$ fluxes \eqref{scaleseparation2}.
    \item Parametric weakly coupled solution with a shrinking dual $s^5$ cycle and scale separation exists, as long as the fluxes respect the following relations
    \begin{equation}
        \mathcal{G} \ll G^2N\quad\text{and}\quad N^3 \mathcal{G}^{-3}\ll G^2\ll\mathcal{G}N^{\frac13}\quad \text{and}\quad G^4\ll N^3\,.
    \end{equation}
    This solution is the T-dual of the classical type IIA background of \cite{Farakos:2025bwf}, albeit slightly more general due to the additional freedom in the flux sector.
    \item Parametric strongly coupled solution, all large cycles and scale separation
    \begin{equation}
       G^6 \ll G^2 N^3 \ll \mathcal{G}\,.
    \end{equation}
    This is a new type I solution in which the $F_7$ flux dominates over $F_3$, as described in \eqref{weakstringcoupling2}. It corresponds to a weakly coupled heterotic solution that will appear soon in \cite{Het}.
\end{itemize}
In these solutions, all moduli are stabilized with the aid of the extra fluxes, as presented in subsection \ref{minimalanisotropic}.

\subsubsection{Masses and dimensions of dual operators}

To calculate the conformal dimensions of the dual CFT for the above construction with the flux configuration in \eqref{F3anisotropic2}, we first compute the derivatives of the scalar potential in \eqref{SUGRAscalarpotential} to obtain the Hessian $V_{IJ}$, and using the moduli space metric, evaluate everything at the vacuum to find the following quantities that only depend on the $G$ and $Q$ flux
\begin{equation}\label{eigen2}
    m^2L^2=\left\{  \frac{8 Q (2 G+3 Q)}{(2 G+Q)^2}, 8, 8, 8, 8, e_1(Q,G), e_2(Q,G), e_3(Q,G)\right\} \,.
\end{equation}
The functions $e_i(Q,G)$ are slightly lengthy expressions. They can be expanded for $G \gg 1$ to reproduce the results from the previous example
\begin{align}
    \e_1(Q,G)&= 48+\frac{35 Q^2}{3 G^2}+\mathcal{O}\left(G^{-3}\right)\,, \\
    \e_2(Q,G)&= -\frac{4 Q}{G}+\frac{22 Q^2}{3 G^2}+\mathcal{O}\left(G^{-3}\right)\,, \\
    \e_3(Q,G)&= \frac{3 Q^2}{G^2}+\mathcal{O}\left(G^{-3}\right)\,.
\end{align}
Note from the expansion of $\e_2(Q,G)= -\frac{4 Q}{G}+\ldots$ that we now have a tachyonic direction that for large $G\gg1$ is however well below the Breitenlohner-Freedman bound of -1. The first value for $m^2L^2$ above becomes in the large $G$ limit $\frac{8 Q (2 G+3 Q)}{(2 G+Q)^2} =\frac{4 Q}{G}+\ldots$ so that two of the previously massless fields have opposite masses squared.

Finally, the conformal dimensions reduce to the previous result for large $G$ but since the masses are now generically flux dependent quantities the dual conformal dimensions are not integers anymore and depend in particular on the flux quanta values $Q$ and $G$.
These results match exactly those of the first isotropic example in type IIA \cite{Farakos:2025bwf}, which is expected since the two theories are dual.

\section{Conclusion}\label{sec:Conclusion}

In this paper, we performed three T-dualities on the classically scale-separated AdS$_3$ solutions found in \cite{Farakos:2025bwf} for massive type IIA flux compactifications on $G_2$-holonomy spaces. We obtain dual type I backgrounds with $G_2$-structure and corresponding non-zero curvature for the internal space. The three T-dualities were performed along directions transverse to an O6-plane canceled by D6-branes in type IIA, producing O9-planes canceled by D9-branes and yielding a type I background. 
The unbounded IIA fluxes that give parametric control map to closed type I fluxes with the same unbounded scaling, yielding parametric behavior in the type I setup, while bounded IIA fluxes map to bounded type I fluxes and ensure tadpole cancellation.
Beyond the dual solution with parametrically small cycles, a flux-scaling analysis yields scale-separated families that are either classical or at strong coupling, with the latter dual to heterotic classical solutions \cite{Het}. The flux configuration inherited from type IIA ensures moduli stabilization and nontrivial tadpole cancellation, and it reproduces on the type I side the type IIA result that in the dual CFT operator conformal dimensions become parametrically integer.

As a future direction, it would be interesting to perform one or two T-dualities starting from the massive type IIA setup, or to use the nilmanifold classification in \cite{VanHemelryck:2025qok}, to generate dual theories with different Betti numbers and test whether scale-separated vacua and the properties above persist.
Another interesting direction is to uplift the type IIA setups to M-theory, following \cite{Cribiori:2021djm}, and analyze the resulting solutions and internal geometry, although passing to massless type IIA introduces O4-planes, which can complicate the uplift.

\section*{Acknowledgments}
We would like to thank Vincent Van Hemelryck for discussions and comments on our results at the EISA Corfu workshop. This work is supported in part by the NSF grant PHY-2210271, the Lehigh University CORE grant with grant ID COREAWD40, and the ERC Starting Grant QGuide101042568 - StG 2021. Z.M. and M.R. acknowledge the support of the Dr. Hyo Sang Lee Graduate Fellowship from the College of Arts and Sciences at Lehigh University. Z.M., M.R., and G.T. would like to thank the Simons Center for Geometry and Physics, where part of this work was completed, for their hospitality.

\appendix

\section{Type II bosonic action}\label{app:typeIIaction}

We will perform the dimensional reduction starting from the 10D bosonic action in Einstein frame. To keep track of conventions, we begin with the 10D bosonic action in the string frame, which takes the following form
\begin{align}\label{StringFrameAction}
    S&=\frac{1}{2\kappa_{10}^2}\int \text{d}^{10}X\sqrt{-G^S}e^{-2\phi}\left(R_{10}+4(\partial\phi)^2
    - \frac{1}{2}\vert H_3\vert^2
    - \frac{1}{4}e^{2\phi}\sum_{q}\vert F_q\vert^2
    \right)\,,
\end{align}
where $2\kappa^2_{10}=(2\pi)^7\alpha^{\prime 4}$.
The RR field strengths $F_q$ depend on the type II theory
\begin{align}
    \text{IIA} \quad &:\quad q=0,2,4,6,8, 10 \\
    \text{IIB} \quad &:\quad q=1,3,5,7,9
\end{align}
For the local objects considered here, O$p$-planes and D$p$-branes, the DBI and Wess-Zumino actions, are given by
\begin{align}\label{sources}
    S_{loc}&=-\mu_{Op/Dp}\int\text{d}^{p+1}X e^{-\phi}\sqrt{-P[G^S]}+\mu_{Op/Dp}\int C_{p+1}\,,
\end{align}
where $P[G^S]$ is the determinant of the pull-back of the ten-dimensional metric $G_{MN}^S$ and $\mu_{Op/Dp}$ denotes the tension and charge of our BPS sources. In particular, we have 
\begin{equation}
    \mu_{Dp} = \frac{1}{(2\pi)^p (\alpha')^{\frac{p+1}{2}}}
    \,,\quad\quad
    \mu_{Op} = -2^{p-5}\mu_{Dp} \,.
\end{equation}

We take the 10D space to be a direct product of two manifolds, $M_{10} = M_{3} \times X_{7}$.  We use capital Latin indices $M, N = 1, \dots, 10$ for the 10D coordinates, Greek letters $\mu, \nu$ for the coordinates of the external 3-dimensional space, and lowercase Latin letters $i, j$ for the internal space indices.

We will reduce the action in Einstein frame, so we perform the following Weyl rescaling 
\begin{equation}\label{WeylRescaling}
    G_{MN}^S=e^{\phi/2}G_{MN} \,.
\end{equation}
The combined bosonic action in Einstein frame is given by 
\begin{equation}\label{EinsteinFrameAction}
\begin{split}
    S_{E}&=\frac{1}{2\kappa_{10}^2}\int \text{d}^{10}X\sqrt{-G}\left(R_{10}-\frac{1}{2}(\partial\phi)^2
    - \frac{1}{2}e^{-\phi}\vert H_3\vert^2
    - \frac{1}{4}e^{\frac{5-q}{2}\phi}\sum_{q}\vert F_q\vert^2
    \right)\, \\
    &+\sum_p \lp -\mu_{Op/Dp}\int\text{d}^{p+1}X \,e^{\frac{p-3}{4}\phi}\,\sqrt{-P[G]} +\mu_{Op/Dp}\int C_{p+1} \rp
    \,.
\end{split}
\end{equation}
We set $\alpha^{\prime} = l_s^2 = (2\pi)^{-2}$ so that $(2\kappa_{10}^2)^{-1} = \mu_{Dp} = 2\pi$. With these conventions, let $b_p$ be the $p$-th Betti number, and let $\Omega_1,\ldots,\Omega_{b_p}$ be a basis for the integer cohomology $H^p(X,\mathbb{Z})$. Then a harmonic $p$-form flux expands as
\begin{equation}
F_p
= (2\pi\sqrt{\alpha^{\prime}})^{p-1}\sum_{i=1}^{b_p}f_p^{i}\,\Omega_i= \sum_{i=1}^{b_p}f_p^{i}\,\Omega_i\,,
\end{equation}
where in the second equality we used $2\pi\sqrt{\alpha'}=1$.

\section{\texorpdfstring{Scalar potentials from dimensional reduction}{}}\label{app:DimensionalReduction}
Reducing the 10D action of \eqref{EinsteinFrameAction} using the metric ansatz \eqref{eq:metric} and \eqref{abrelation}, we obtain the 3D Einstein-frame action
\begin{equation}
    S_3 = 4\pi \int \sqrt{-g} \lp \frac12 R - G_{IJ} \partial_\mu \varphi^I \partial^\mu \varphi^J - V(\varphi^I) \rp\,,
\end{equation}
where we have pulled out an additional overall factor of 2 before defining the moduli space metric $G_{IJ}$ and the scalar potential $V(\varphi^I)$. In 3D supergravity the scalar potential can be written in terms of a superpotential $P(\varphi^I)$ and the moduli space metric via
\begin{equation}
    V= G^{IJ} P_I P_J - 4 P^2\,,
\end{equation}
with $P_I = \partial_{\varphi^I} P$.

The potential energy contributions, that arise after the compactification of the ten-dimensional action considering the reduction ansatz in equation \eqref{eq:metric}, are the following 
\be
\begin{aligned}
    V_R &= -\frac12 \tilde R^{(7)} \, e^{-16\beta \upsilon} \, ,  \\
    V_{H_3} &= \frac{1}{4}\vert \tilde{H}_3 \vert^2  e^{-\phi} e^{-20\beta \upsilon} \, ,  \\
    V_{F_q} &= \frac{1}{8} \vert \tilde{F}_q \vert^2 \, e^{\frac{5-q}{2}\phi} e^{-2(7+q)\beta \upsilon} \, ,  \\
    V_{Dp/Op} &= \sum_p \frac12 \frac{\mu_{Op/Dp}}{2\pi}  \, e^{\frac{p-3}{4}\phi} \,  e^{(p-23)\beta \upsilon}\label{DPOP} \,.
\end{aligned}
\ee

\subsection{Scalar potential from type IIB}

We focus on type IIB with an O5/O9 orientifold projection and only $F_3$ and $F_7$ fluxes. The relevant contributions to the scalar potential obtained after dimensional reduction take the form
\begin{equation}\label{scalarpotentialIIB}
\begin{split}
    V&=V_R+V_{F_3}+V_{F_7}+V_{D5/O5} \\
    &=-\mathcal{R}_0(\tilde{s}^i)e^{-16\beta\upsilon}
    +\mathcal{F}_3(\tilde{s}^i)e^{-20\beta\upsilon+\phi}
    +\mathcal{F}_7e^{-28\beta\upsilon-\phi}
    +\mathcal{T}_5(\tilde{s}^i)e^{-18\beta\upsilon+\frac{\phi}{2}} \,,
\end{split}
\end{equation}
where we have canceled the tension of the O9-plane by adding D9-branes. 
The calligraphic functions encode the metric deformations of the internal space and the flux quanta. They can be computed explicitly
\begin{align}
    \mathcal{R}_0(\tilde{s}^i)&=\frac12 \tilde{R}_7=\frac{1}{16}\left(\sum_{i,j}\tilde{s}^i\mathcal{M}_{ij}\tilde{s}^j\right)^2-\frac{1}{4}\sum_j \lp \sum_{i}\tilde{s}^i\mathcal{M}_{ij}\rp^2 \lp \tilde{s}^j\rp^2 \,,\label{fun1IIB}\\
    \mathcal{F}_7&=\frac{\mathcal{G}^2}{4}\,, \label{fun2IIB}\\
    \mathcal{F}_3(\tilde{s}^i)&=\frac{1}{4}\int_7\tilde{\star}F_3\wedge F_3=\frac{1}{4}\sum_{i=1}^7\left(\frac{f^i}{\tilde{s}^i}\right)^2\,, \label{fun3IIB}\\
    \mathcal{T}_5(\tilde{s}^i)&=\frac12 \sum_{k,l}f^l\mathcal{M}_{lk}\tilde{s}^k \,. \label{fun4IIB}
\end{align}
Note that $F_7=\star_{10} F_3$, where $\star_{10}$ is the 10D Hodge star and this leads to an effective doubling of their respective contributions. We also used the tadpole cancellation condition to express the source contributions from Op/Dp sources in terms of the fluxes.
As discussed in the main text this scalar potential follows from 
\begin{equation}
    P=\frac{1}{4 \text{vol}(X)^2} \int_X \lp e^{-\frac{\phi}{2}} F_7 +e^{\frac{\phi}{2}} \star \hat{\Phi} \w F_3 +\frac12 \hat{\Phi}\wedge d\hat{\Phi} \rp \,.
\end{equation}

\section{\texorpdfstring{The explicit dual fluxes and sources}{}}\label{Dualfieldsexplicit}

We perform the three T-dualities along the directions $y^2, y^3,$ and $y^5$ on the non-minimal flux in equation \eqref{F4nonminimal} (which also includes the minimal case in equation  \eqref{F4minimal} if set $G=Q=0$). Acting from the left in the order $T_{y^5}T_{y^3}T_{y^2}$, and writing the result collectively and explicitly, we have 
\begin{align}\label{fluxes1example1}
    f^{(1)}_4\Psi_1= +(Q+G)d\eta^{3456}\quad&\leftrightarrow\quad -(Q+G)d\eta^{246}=-(Q+G)\hat{\Phi}_7 \,,\\
    f^{(2)}_4\Psi_2= -(-N)d\eta^{1256}\quad&\leftrightarrow\quad -(-N)d\eta^{136}= -(-N)\hat{\Phi}_4 \,,\\
    f^{(3)}_4\Psi_3= -(Q+G)d\eta^{1234}\quad&\leftrightarrow\quad -(Q+G)d\eta^{145}= -(Q+G)\hat{\Phi}_6 \,,\\
    f^{(4)}_4\Psi_4= +(Q-2G)d\eta^{2457}\quad&\leftrightarrow\quad +(Q-2G)d\eta^{347}= -(Q-2G)\hat{\Phi}_2 \,,\\
    f^{(5)}_4\Psi_5= -(-N)d\eta^{1467}\quad&\leftrightarrow\quad +(-N)d\eta^{1234567} \,, \\
    f^{(6)}_4\Psi_6= +(-N)d\eta^{2367}\quad&\leftrightarrow\quad +(-N)d\eta^{567}= -(-N)\hat{\Phi}_3 \,,\\
    f^{(7)}_4\Psi_7= +(-nN)d\eta^{1357}\quad&\leftrightarrow\quad -(-nN)d\eta^{127}= -(-nN)\hat{\Phi}_1 \,,\\
    m_0\quad&\leftrightarrow\quad -m_0d\eta^{235}=m_0\hat{\Phi}_5 \,.
\end{align}
In the above expressions we have not simplified the signs to shows how they arise from two different places, e.g., $f^{(2)}_4\Psi_2= N\,d\eta^{1256}$ because $f^{(2)}_4 = (-N)$ and $\Psi_2=-d\eta^{1256}$.
Collecting these terms, we write them as the expansion \eqref{basisF7F3} in the chosen basis
\begin{equation}
\begin{split}
    F_3=&+nN\hat{\Phi}_1+N\hat{\Phi}_3+N\hat{\Phi}_4+m_0\hat{\Phi}_5\\
    &-(Q+G)\hat{\Phi}_6-(Q+G)\hat{\Phi}_7-(Q-2G)\hat{\Phi}_2 \,.
\end{split}
\end{equation}
In order to find AdS solutions we find that the $N, Q,$ and $G$ fluxes must have the opposite sign to the $m_0$ flux. To simplify explicit expression we will therefore replace $m_0 \to -\tilde{m}_0$. Since the theory is invariant under a simultaneous flip of the signs of all RR and NSNS fluxes, we can then chose all flux quanta to be positive
\begin{equation}
    N>0\,,\quad n>0\,,\quad G>0 \,,\quad Q>0 \,,\quad \tilde{m}_0>0\,.
\end{equation}

Applying the three T-dualities in equation \eqref{Tdualities1} to the fifth $F_4$ component gives
\begin{equation}
    F_7 = -N d\eta^{1}\wedge\dots\wedge d\eta^{7}\,.
\end{equation}
In what follows we keep this flux general and denote it by $\mathcal{G}$, i.e. $F_7=-\mathcal{G}\,d\eta^{1\dots 7}$. Since $\mathcal{G}$ is not constrained by any tadpole (on either the type I or IIA side), its amplitude can be different from the other $N$ fluxes, and this generalizes the previous solutions. Also, the sign is a priori arbitrary, $sign(\mathcal{G})=\pm 1$. Supersymmetric stabilization fixes the sign and in our conventions it requires $\mathcal{G}>0$. 

Apart from the fluxes, we also write down the currents of the O6-planes that are canceled by fluxes and transform into currents of O5-planes on the type I side:
\begin{align}
    \quad J_{3,\alpha}= \Phi_{3} = -dy^{567}\quad&\leftrightarrow\quad J_{4,}=+d\eta^{2367}=+\Psi_{6} \,, \label{current1}\\
    \quad J_{3,\alpha\beta}= \Phi_{1}= +dy^{127}\quad&\leftrightarrow\quad J_{4,}= +d\eta^{1357}=+\Psi_{7} \,, \label{current2} \\
    J_{3,\gamma\alpha}= \Phi_{4}= +dy^{136}\quad&\leftrightarrow\quad J_{4,}=-d\eta^{1256}= +\Psi_{2} \,, \label{current3} \\
    J_{7}=+d\eta^{1234567}\quad&\leftrightarrow\quad J_{4,\alpha\beta\gamma}=+d\eta^{1467}= -\Psi_{5} \,. \label{current4}
\end{align}
In the table below, we map the O-planes and D-branes in type IIA to the corresponding O-planes and D-branes obtained after performing the three T-dualities along $y^2,y^3,y^5$:
\begin{align}\label{IIAIsources}
\begin{pmatrix} 
&{\rm O}2/{\rm D}2:\quad & - & - & - & - & - & - & -  \\
&{\rm O}6_{\alpha}:\quad & \times & \times & \times & \times & - & - & -  \\
&{\rm O}6_{\beta}/{\rm D}6_{\beta} :\quad &\times & \times & - & - & \times & \times & -  \\
&{\rm O}6_{\gamma}/{\rm D}6_{\gamma}:\quad &\times & - & \times & - & \times & -& \times  \\
&{\rm O}6_{\alpha\beta}:\quad & - & - & \times & \times & \times & \times & -  \\ 
&{\rm O}6_{\beta\gamma} :\quad & - & \times & \times & - & - & \times & \times  \\ 
&{\rm O}6_{\gamma\alpha}/{\rm D}6_{\gamma\alpha} :\quad &- & \times & - & \times & \times & - & \times  \\ 
&{\rm O}6_{\alpha\beta\gamma}/{\rm D}6_{\alpha\beta\gamma}:\quad &\times & - & - & \times & - & \times & \times 
\end{pmatrix}
\rightarrow
\begin{pmatrix} 
&{\rm O}5^{\prime}_{\alpha\beta\gamma}/{\rm D}5^{\prime}_{\alpha\beta\gamma}:\quad & - & \times & \times & - & \times & - & -  \\
&{\rm O}5_{\beta\gamma}:\quad & \times & - & - & \times & \times & - & -  \\
&{\rm O}5_{\gamma\alpha}/{\rm D}5_{\gamma\alpha} :\quad &\times & - & \times & - & - & \times & -  \\
&{\rm O}5_{\alpha\beta}/{\rm D}5_{\alpha\beta}:\quad &\times & \times & - & - & - & - & \times  \\
&{\rm O}5_{\gamma}:\quad & - & \times & - & \times & - & \times & -  \\ 
&{\rm O}5_{\alpha} :\quad & - & - & - & - & \times & \times & \times  \\ 
&{\rm O}5_{\beta}/{\rm D}5_{\beta} :\quad & - & - & \times & \times & - & - & \times  \\ 
&{\rm O}9/{\rm D}9:\quad &\times & \times & \times & \times & \times & \times & \times 
\end{pmatrix} \, . 
\end{align}

\pagebreak

\bibliographystyle{JHEP}
\bibliography{refs}

\end{document}